\documentclass[conference]{IEEEtran}
\IEEEoverridecommandlockouts
\usepackage{cite}
\usepackage{amsmath,amssymb,amsfonts}
\usepackage{algorithmic}
\usepackage{graphicx}
\usepackage{textcomp}
\usepackage{xcolor}

\usepackage{amsthm}
\usepackage{enumitem}
\usepackage{hyperref}
\usepackage{bm}

\def\BibTeX{{\rm B\kern-.05em{\sc i\kern-.025em b}\kern-.08em
    T\kern-.1667em\lower.7ex\hbox{E}\kern-.125emX}}
\begin{document}

\newcommand{\p}{\mathbb{P}}
\renewcommand{\d}{\delta}
\renewcommand{\a}{\alpha}
\newcommand{\indep}{\perp\!\!\!\perp}
\newcommand{\simiid}{\overset{\text{i.i.d.}}{\sim}}
\newcommand{\var}{\text{Var}}
\newcommand{\sign}{\text{sign}} 
\newcommand{\support}{\textrm{support}}
\newcommand{\Nuniq}{N_{\text{unique}}}
\newcommand{\Nmax}{N_{\text{max}}}
\newcommand{\Nnz}{N_{\text{nz}}}
\newcommand{\ah}{\hat{\alpha}}

\title{SAIVE: Selecting AI Valuable Entities
\thanks{Research was sponsored by the Department of the Air Force Artificial Intelligence Accelerator and was accomplished under Cooperative Agreement Number FA8750-19-2-1000. The views and conclusions contained in this document are those of the authors and should not be interpreted as representing the official policies, either expressed or implied, of the Department of the Air Force or the U.S. Government. The U.S. Government is authorized to reproduce and distribute reprints for Government purposes notwithstanding any copyright notation herein.}
}

\author{\IEEEauthorblockN{
Inna Voloshchuk, Hayden Jananthan, Jeremy Kepner
\\
\IEEEauthorblockA{
MIT
}}}

\IEEEpubid{\makebox[\columnwidth]{979-8-3195-2794-3/26/\$31.00 \copyright2026 IEEE \hfill} \hspace{\columnsep}\makebox[\columnwidth]{ }}

\maketitle
\IEEEpubidadjcol

\begin{abstract}

Data lakes store large amounts of telemetry, with logs from network sensors, hosts, and applications containing possibly hundreds of fields for every event. Large enterprises are then left with data lakes that cannot be analyzed efficiently with AI. Aggregate analysis looks at persistent shifts in behavior over time. Many of the fields and columns in data lakes are not useful as they do not contain information that is sufficiently diverse or concentrated to support AI analysis. SAIVE is a simple method for examining a few rows in a large table and applies a histogram of histograms filtering criterion to select the fields that for AI analysis is more likely to yield useful results. This paper provides a principled foundation for the SAIVE heuristics by assuming of a Zipf-Mandelbrot power-law distribution of the underlying data.  Constraining the Zipf-Mandelbrot exponent alpha to a reasonable range provides a a practical, cheap, expert-free filter for selecting AI valuable entities in large data sets.
\end{abstract}


\section{Introduction}

 Data lakes store large of amounts of data that is routinely unanalyzed by AI because the volume is prohibitive.  This is particularly relevant to cybersecurity where telemetry is a crucial part of the defensive security of various systems, and is collected over many hosts and long periods of time~\cite{vielberth2020security, shahjee2022integrated, du2017deeplog, zhang2019robust, scarfone2008computer}. The resulting data volume, velocity, and variety~\cite{laney20013D} make analyzing entries and events with AI complex when the goal is to find subtle changes in behavior over long durations and large groups of actors. Thus, there is interest in reducing the total amount of data to analyze using cheap-to-compute statistics, bringing more value to organizations.
 
Prior work on scalable analysis has developed practical heuristics for choosing useful fields for further analysis \cite{Vijay2014BigData, schofield2026CyberLog}.  These heuristics  form the basis of SAIVE which can be summarized as follows.  Given a column of data with $N$ rows, let
\begin{enumerate}
    \item $N_{\rm nonempty}$: number of nonempty entries containing at least one observed value
    \item $N_{\rm unique}$: number of distinct values
    \item $N_{\rm max}$: count of the most popular value
\end{enumerate}
Empirically it has been observed the columns supporting additional analysis adhere to the following criteria
\begin{eqnarray*}
& N_{\rm nonempty} & \sim N \\
1 \ll & N_{\rm unique} & \ll N \\
1 \ll & N_{\rm max} &  \ll N
\end{eqnarray*}
 

The goal of this work is to develop a more principled basis for SAIVE heuristics that will lead to automated procedures for selecting useful data.
Previous work has found that many relevant quantities are power law and can be modeled using a modified Zipf-Mandelbrot distribution \cite{kepner202210, barabasi2016network}.
Using a power law model as a base assumption,
constraining the power-law slope parameter $\a$ to a reasonable range is one approach for putting the SAIVE heuristics on a firmer mathematical footing.

The rest of the paper is organized as follows. Section~\ref{sec:arr} introduces the mathematical preliminaries for representing tables and columns as associative arrays \cite{Vijay2014BigData, kepner2018mathematics}. In Section~\ref{sec:ZM}, we define the Zipf-Mandelbrot model and derive the closed-form approximation of $\a$. Section~\ref{sec:exp} describes the experiment on synthetic data before a discussion of the results and the limitations in Section~\ref{sec:res}, and Section~\ref{sec:conc} concludes.

\section{Associative Arrays}\label{sec:arr}

Let ${\bf T}$ be an associative array (table) with $N$ rows and $N_{\rm col}$ columns, with values $v$ from alphabet $V$. The value of row $r$ and column $c$ is
\[
  {\bf T}(r,c) = v \in V.
\]
For a given column $c$, let the number of unique values be $N^c_{\rm unique}$.
Construct an $N{\times}N^c_{\rm unique}$ associative array ${\bf A}_c$ for each column
\[
{\bf A}_c(r, v) = \begin{cases}
    1, & {\bf T}(r,c)=v\\
    0, & {\rm otherwise}
\end{cases}
\]
Each row of ${\bf A}_c$ contains exactly one nonzero entry.
Summing over the rows of ${\bf A}_c$ results in an $1{\times}N^c_{\rm unique}$ associative array where each entry is the histogram of a particular value in the column $c$
\[
{\bf d}_c = {\bf 1}^\top {\bf A}_c, ~~~~ {\bf d}_c(v) = \sum_{r=1}^N {\bf A}_c(r,v)
\] 
where ${\bf 1}$ as $N{\times}1$ associative array of all 1's and
\[
\sum_{v=1}^{N^c_{\rm unique}} {\bf d}_c(v) = N , ~~~~ \Nmax^c = \max {\bf d}_c
\]

Similarly, an additional $N^c_{\rm unique}{\times}\Nmax^c$  associative array ${\bf D}_c$ can be constructed from ${\bf d}_c$
\[
{\bf D}_c(v,d) = \begin{cases}
    1, & {\bf d}_c(v) = d \\ 
    0, & {\rm otherwise}
\end{cases}
\]
Summing over the rows of ${\bf D}_c$ results in an $1{\times}\Nmax^c$ \emph{matrix} (or row vector) ${\bf n}_c$, where each entry is the histogram of values $v$ with the same count $d$ (i.e., the histogram of histograms)
\[
{\bf n}_c = {\bf 1}^\top {\bf D}_c ,  ~~~~ {\bf n}_c(d) = \sum_{v=1}^{N^c_{\rm unique}} {\bf D}_c(v,d)
\]
with
\[
\sum_{d=1}^{\Nmax^c} {\bf n}_c(d) ~ d = N , ~~~~ N^c_{\rm unique} = \mathbf{1^\top} {\bf n}_c
\]

The vector ${\bf n}_c$ can be interpreted as the histogram of value frequencies, where ${\bf n}_c(d)$ counts how many distinct values of $v$ occurred exactly $d$ times.  Normalizing ${\bf n}_c$ produces the corresponding probability $p_c(d)$ that can be modeled with a Zipf-Mandelbrot distribution. For the remainder of the paper, the per column $c$ notation  will be dropped for brevity with all of following analysis is done column by column for the table ${\bf T}$.

\begin{figure*}
\center{\includegraphics[width=2.0\columnwidth]{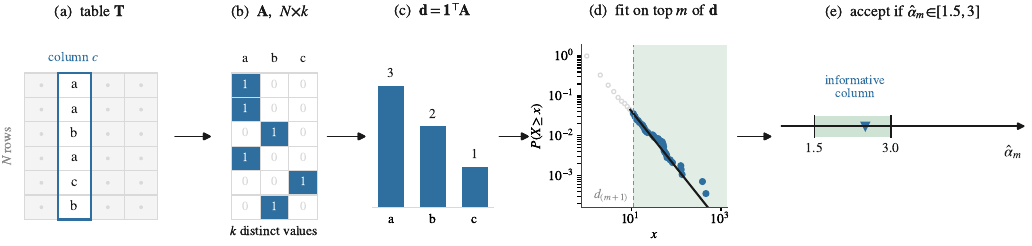}}
      	\caption{{\bf Overview of the column model.} A table of categorical data can be represented using associative arrays if values are assumed meaningless. The alpha parameter of the Zipf-Mandelbrot distribution is estimated on the frequency of each categorical value. Previous work on real-world datasets has shown that typical values of this parameter fall in between $1.5$ and $3$, and thus is one descriptor of a column that could be informative. }
      	\label{fig:overview}
\end{figure*}

\section{Zipf-Mandelbrot Distribution}\label{sec:ZM}

A useful column should have enough repeated information to support aggregate analysis yet enough diversity to discern behaviors~\cite{schofield2026CyberLog}. Thus, the object studied is the distribution of value frequencies. Many quantities of interest exhibit these qualities, and it has been shown that these quantities follow a modified Zipf-Mandelbrot distribution \cite{kepner202210}. Note that this distribution is applied to the actual count values, as opposed to the rankings of categories based on frequency, as many power-law models are usually applied. Given a table ${\bf T}$, the associative arrays ${\bf A}$, ${\bf D}$, ${\bf d}$, and vector ${\bf n}$ for each column are constructed as described in~\ref{sec:arr}.
The value frequencies, $d$, are modeled as independent and identically distributed samples from a Zipf-Mandelbrot distribution on $\{1, 2, \ldots, N_{\rm max}\}$
\begin{equation*}\label{eq:ZM_pmf}
    p(d) = \frac{(d+\delta)^{-\a}}{H(\Nmax, \d, \a)}
\end{equation*}
where 
\[
H(N_{\rm max}, \d, \a) = \sum_{i=d}^{N_{\rm max}} \frac{1}{(d +\d)^\a}
\]
Specifically, for some $\alpha$ and $\delta$
\[
x_1, \dots, x_k \overset{\text{i.i.d.}}{\sim} \text{Zipf-Mandelbrot}(\a, \d)
\]

The $k$ samples taken
represent the number of distinct values seen. The exponent $\a$ represents the slope of the distribution. For a large $\a$, larges values of $x_i$ are rare, while for a small $\a$ near $1$,  $x_i$ can  be very large. The Mandelbrot offset $\d$ helps fit the head of the distribution, at small sampled values. It is important to note that in heavy tailed distributions, the mean and variance may diverge and be undefined \cite{nair2022fundamentals}. 

The empirical work of~\cite{kepner202210} finds that many quantities fit a modified Zipf-Mandelbrot distribution with an $\a$ in the range of $1.5$ to $3.$ Since $\a$ is the main descriptor of a Zipf-Mandelbrot distribution, the estimation of $\a$ can be used as a potential criterion to filter out  columns which may not be amenable to AI analysis and enables removing columns which have an estimated $\a$ outside of this range.

\subsection{Approximating $\a$}
The natural way of estimating a parameter from a distribution given $k$ i.i.d. draws is through maximum likelihood estimation. Under the Zipf-Mandelbrot model, the log-likelihood of the sample is
\begin{align*}
    \ell(\alpha) &= \sum_{i=1}^k \ln p(x_i) \\
                 &= -\a \sum_{i=1}^{k} \ln(x_i+\d) - k \ln H(\Nmax, \d, \a)
\end{align*}

The normalization constant $H(\Nmax, \d, \a)$ prevents a closed form solution for $\a$ in the above log-likelihood equation. Thus, an approximation of the discrete likelihood estimator for $\a$ is obtained by replacing the normalization sum with an integral and taking the untruncated limit of $\Nmax \rightarrow \infty$, as given in~\cite{clauset2009power}. For this estimator, we keep the Mandelbrot offset of $\d$, but the reader can find more details on the similar algebra in~\cite{clauset2009power}. 

For $\a> 1$, this gives the approximate estimator
\begin{align*}
    \ah &= 1+\frac{k}{\sum^k_{i=1} \ln(\frac{x_i+\d}{1/2+\d})}
\end{align*}

Even when the tail of a distribution follows a power law, the smallest observations may not~\cite{nair2022fundamentals}. Additionally, the lower end of a discrete distribution is strongly affected by the offset parameter $\d$, which is not known a priori. When only getting a sample, the smallest observations, where $\d$ matters the most, is also often the most unreliable. To avoid this, it is natural to only fit the upper tail that does exhibit the power law behavior, which is the idea of a Hill estimator~\cite{hill1975simple}. Without loss of generality we can reorder the samples from largest to smallest so that
\[
x_1 \geq x_2 \geq \cdots \geq x_k 
\]
Using the largest $m$ observations, the corresponding estimator has the form 
\begin{equation}\label{eq:ahm}
\ah_m = 1 + \frac{m}{\sum_{i=1}^m \log \left( \frac{x_i + \d}{x_{m+1} + \d -\frac12} \right)} 
\end{equation}

There is a bit of a difficulty in picking one universal number $m$ to use in the equation of the approximate $\a$ estimate for all the columns in a table. The usual procedure requires one to create Hill plots for different values of cutoff $m$, with a statistician to then pick the appropriate value of $m$, which can be prone to error. Picking an appropriate $m$ would need to be repeated for every column. This procedure is not feasible and goes against the very motivation of the problem where the goal was to instead have an automatic criterion that would be applied to columns without expert intervention. Thus, a fixed $m$ is used for all columns.

Using estimation of the $\a$ parameter as a proxy for a column being informative, columns that do not exhibit power-law behavior can be removed, and are therefore not of interest, reducing the amount of columns to analyze in downstream tasks. The acceptable range of alpha is in $[1.5, 3]$, and this interval is an empirical design choice motivated by the results of~\cite{kepner202210}. For a full overview of the progression from a table of categorical values to the formula of alpha, refer to Figure~\ref{fig:overview}. 

\section{Synthetic Data Experiment}\label{sec:exp}

\begin{figure*}
\center{\includegraphics[width=2\columnwidth]{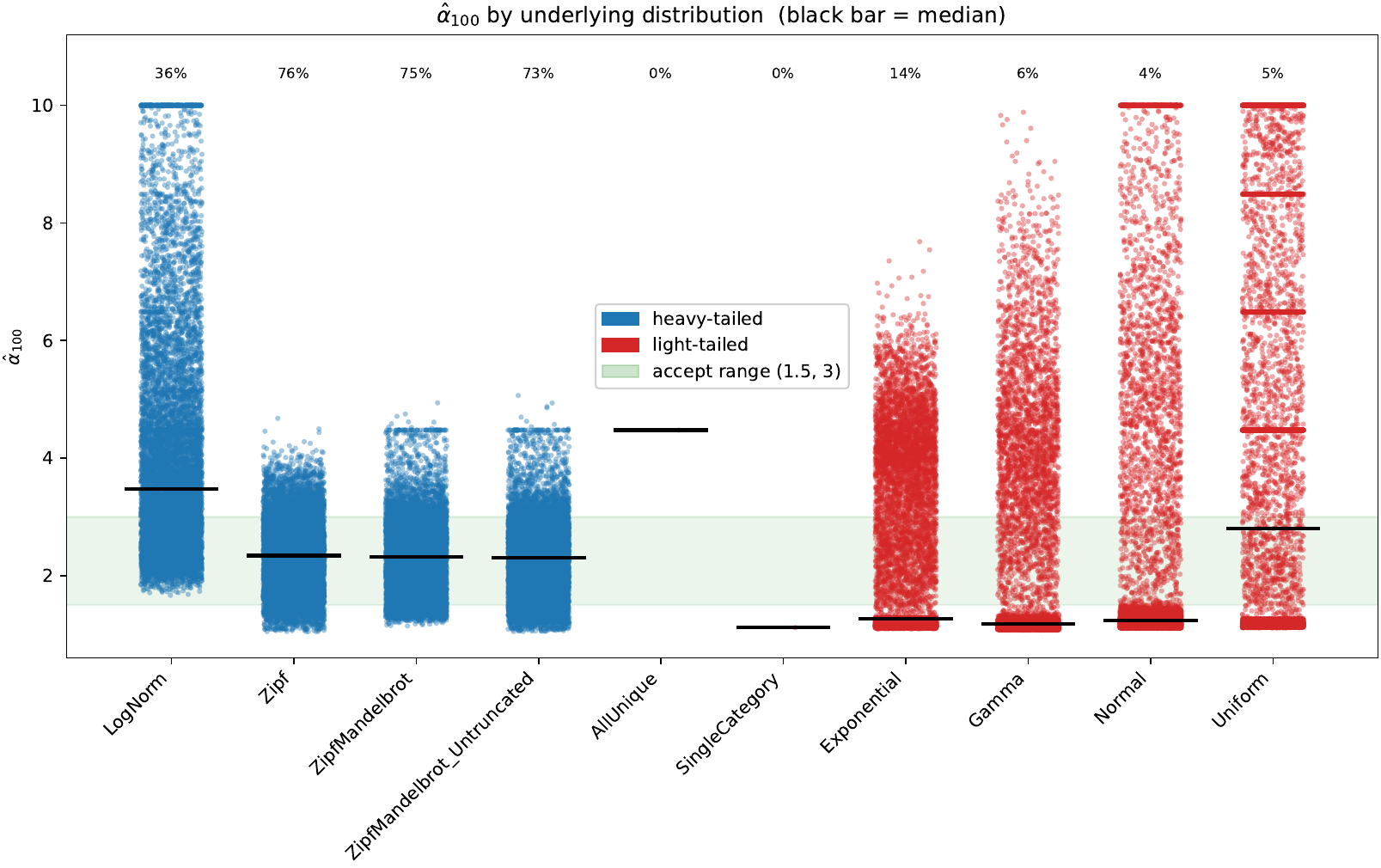}}
      	\caption{{\bf Result of Synthetic Data Set Experiment.} Percentages above each column give the fraction of trials accepted by the $\ah_m \in [1.5,3]$ filter. In this experiment, $N=8192, m=100, \d = 1,$ and each distribution family gets sampled from $10,000$ times. The values of $\ah_m$ for each sample are cut off at $10$ to better display the figure.}
      	\label{fig:result}
\end{figure*}

To test how well a filter based on the estimator Equation~\ref{eq:ahm} removes non-power-law columns, we apply the filter onto synthetic data with controlled generating processes, even those which do not have a power-law exponent to estimate. 

Each trial creates a column of data by drawing a value from a chosen distribution until the total sum of draws reaches $N$. This method of drawing count values creates the associative array ${\bf d}$ for that column, which is all that is needed to then evaluate an $\ah_m.$ 

For this experiment, we chose $N=8192,$ as this matches the row count of the cyber log tables in~\cite{schofield2026CyberLog}. Additionally, the evaluation of $\ah_m$ for each column requires specifying a value of $\d$. Although there are methods for estimating this value for Zipf-Mandelbrot distributions, because this procedure applies a formula onto data that is known to have values that will not follow this particular distribution, a value of $\d=1$ is chosen. This is chosen to separate the edge cases of a column away from the acceptable range. See that $\d$ must be greater than approximately $0.27$ to ensure that a column containing all unique values, where ${\bf d}  = [1 \, \dots \, 1],$ has an $\ah_m$ value greater than the necessary threshold of $3.$

Different distributions are sampled to simulate different possible generating processes. For heavy-tailed generating distributions, we use the Zipf distribution, the Zipf-Mandelbrot distribution with max value $N$, an untruncated version of the Zipf-Mandelbrot where $N=\infty$, and the log normal distribution. For light tailed distributions, we use the normal, exponential, gamma, and uniform. There are also the deterministic edge cases of a column consisting of all unique values, where all samples are 1, and of a column consisting of a single value, where then the frequency value is always $N$. These distributions were chosen as they were common distributions with different ranges of behavior. 

Each of these distributions has their own parameters, and the ranges of the parameters reflect how much is assumed about each distribution family. For the light-tailed and log normal distributions, the parameters are sampled log uniformly over a large range, from $1$ to over $N$, to allow the experiment to cover several orders of magnitude. This is in an effort to make less assumptions on the possible generating distribution or scale possible for an unknown column, as there is no prior knowledge to constrain the parameters on these distributions. For the power-law distributions, the scale parameter $\a$ is sampled uniformly from the range $1.25$ and $3.5$ and offset $\d$ between $-1$ and $3.5$. This is a narrower range because the network quantities that should be retained are expected to be inside these bounds. Since $\ah_m$ will work well in estimating the scale parameter, choosing from an alpha much outside of this range would commonly be filtered out by the criterion.

The experiment uses a tail cutoff of the top $m=100$ samples, which keeps computation focused on the largest categories seen if there were many samples taken. The appropriate value of $m$ is, as discussed previously, different for each sample, or might not even exist for some distributions. A possible avenue to expand this work would be to explore how changing the value of $m$ changes the performance of the filtering criterion.

For the synthetic data generation, repeat the following $10,000$ times per distribution: 
\begin{enumerate}
    \item Draw the parameters of the distribution at random from the ranges above. 
    \item Sample counts until total count of $N$ is reached, creating associative array ${\bf d}$. 
    \item Calculate the approximated $\ah_m$ using Equation~\ref{eq:ahm} on the $m$ largest samples.
\end{enumerate}

This produces $10,000$ values of $\ah_m$ per distribution, which are summarized in Figure~\ref{fig:result}. The estimator we filter on is a version of the approximation in~\cite{clauset2009power}, keeping a Mandelbrot offset $\d$ and fitting only the largest observations through order statistics. 

\section{Results}\label{sec:res}

The results of the synthetic dataset support the idea that the approximate estimator behaves differently on power-law and non-power-law columns, and a rule on $\ah_m$ can help separate the two.

For power-law distributions like the Zipf and Zipf-Mandelbrot, the estimated exponent works well as it is a meaningful descriptor for the distribution. On the contrary, for the heavy-tailed log normal distribution, the large spread of calculated values occurs because it is not a power-law distribution.

The light-tailed generated values also show a wide spread of $\ah_m,$ which reflect the large parameter ranges that are sampled for these distributions. Their mass clusters near $\ah_m \approx 1$, which occurs when the scale of sampled values from these distributions is large, causing the number of samples taken before the threshold of $N$ is reached to be small. The value of $\ah_m$ is restricted to be under $10$ to make Figure~\ref{fig:result} more focused, and $3\%$ of columns have calculated $\ah_m$ values larger than this that were truncated. The filter of $\ah_m \in [1.5, 3]$ drops $92.6\%$ of the light-tailed columns. The two edge cases are also always dropped, because of the choice of $\d$ for the all unique case, as discussed previously. Together, these results may suggest that a filter on $\ah_m$ can remove a large fraction of columns that do not represent network quantities of interest. 

It is important to note that the estimator is derived using an i.i.d. assumption on the entries of ${\bf d}$, or the drawn values, but in a real table, these frequencies of values must satisfy $\mathbf{1^\top}{\bf d}  = N$. Conditioning on the total number of records in a table makes the frequencies of values dependent. The i.i.d. assumption simplifies the derivation and may match the data well enough for the filter to separate, but it is an approximation to an exact model of a column. The method of Poissonization can be used to remove this dependency~\cite{evert2004ZM}. 

Additionally, the synthetic experiment shows that the statistic behaves as expected on controlled distributions, but it is not yet tested on real cyber log data. An important next step would be to test how the filtering criterion of alpha in Equation~\ref{eq:ahm} would work on real world data, how many fields get discarded, and if the remaining fields are useful information for aggregate analysis. 

\section{Conclusion}\label{sec:conc}
Large tables and data lakes containing cyber log data often require a domain expert to analyze and deem particular columns of values as informative for further aggregate analysis. Under the assumption that column names and value meanings are unknown, these tables can be represented as associative arrays over different categories and their associated frequencies. 

There is often particular interest in network quantities for aggregate analysis to see persistent shifts in behavior in data. For this, we used a Zipf-Mandelbrot model to describe a desirable distribution of value frequencies, and derived a closed form approximation of the exponent alpha. The resulting estimator needs no expert-tuned cutoff, and can be applied on every column of a table and evaluated on the filter of $\ah_m \in [1.5, 3],$ discarding any columns that do not fall in this range. 

Although the results are on controlled distributions, calculation of the approximate alpha formula can be a way to remove non-informative data using cheap-to-compute statistics. This reduces the number of fields necessary to analyze in further stages, lowering man hours and compute necessary to sift through Big Data and extract insights from data lakes. 

\section*{Acknowledgment}
The authors wish to acknowledge the following individuals for their contributions and support: J. Kelner, A. Wierman, L. Anderson, W. Arcand, D. Bestor, W. Bergeron, B. Bond, C. Byun, A. Bonn, D. Burrill, V. Gadepally, J. Gottschalk, T. Hardjono, M. Houle, M. Hubbell, M. Jones, P. Luszczek, P. Michaleas, L. Milechin, J. Mullen, C. Leiserson, C. Milner, S. Mohindra,  A. Pentland, A. Prout, C. Prothmann, A. Reuther, A. Rosa, D. Rus, D. Ross, J. Ross, M. Sherman, S. Van Broekhoven, M. Weems, J. Wilkinson, C. Yee.

\bibliographystyle{ieeetr}
\bibliography{PowerLawCyberLogs}

\end{document}